\documentclass{article}

\usepackage{arxiv}

\usepackage[utf8]{inputenc} 
\usepackage[T1]{fontenc}    
\usepackage{hyperref}       
\usepackage{url}            
\usepackage{booktabs}       
\usepackage{amsfonts}       
\usepackage{nicefrac}       
\usepackage{microtype}      
\usepackage{cleveref}       
\usepackage{graphicx}
\usepackage[numbers]{natbib}
\usepackage{doi}
\usepackage{multirow}
\usepackage{array} 
\usepackage{placeins} 

\title{HISTAI: An Open-Source, Large-Scale Whole Slide Image Dataset for Computational Pathology}
\date{}

\usepackage{authblk}

\author[1]{Dmitry Nechaev\thanks{\texttt{dmitry@hist.ai}}}
\author[1]{Alexey Pchelnikov\thanks{\texttt{alex@hist.ai}}}
\author[1]{Ekaterina Ivanova\thanks{\texttt{kate@hist.ai}}}
\affil[1]{HistAI}

\renewcommand{\headeright}{HISTAI: An Open-Source, Large-Scale WSI Dataset}
\renewcommand{\undertitle}{}
\renewcommand{\shorttitle}{}

\begin{document}
\maketitle

\begin{abstract}
Recent advancements in Digital Pathology (DP), particularly through artificial intelligence and Foundation Models, have underscored the importance of large-scale, diverse, and richly annotated datasets. Despite their critical role, publicly available Whole Slide Image (WSI) datasets often lack sufficient scale, tissue diversity, and comprehensive clinical metadata, limiting the robustness and generalizability of AI models. In response, we introduce the HISTAI dataset, a large, multimodal, open-access WSI collection comprising over 60,000 slides from various tissue types. Each case in the HISTAI dataset is accompanied by extensive clinical metadata, including diagnosis, demographic information, detailed pathological annotations, and standardized diagnostic coding. The dataset aims to fill gaps identified in existing resources, promoting innovation, reproducibility, and the development of clinically relevant computational pathology solutions. The dataset can be accessed at \href{github.com/HistAI/HISTAI}{https://github.com/HistAI/HISTAI}.
\end{abstract}

\section{Introduction}

In recent years, the field of Digital Pathology (DP) has experienced rapid advancement driven largely by progress in artificial intelligence, particularly with the emergence of Foundation Models. These large-scale models, trained on extensive datasets comprising billions of images, have established robust feature extraction capabilities that significantly boost performance on downstream tasks such as classification, segmentation, and diagnostic prediction. However, the effectiveness and generalizability of these AI models are fundamentally limited by the quantity, diversity, and accessibility of the datasets used during training.

Despite the critical role that Whole Slide Images (WSIs) play in computational pathology, publicly available datasets remain scarce compared to other computer vision domains. The majority of existing WSI datasets, while instrumental in setting benchmarks and facilitating supervised learning, are often limited in scale, diversity of tissue types, staining methods, and clinical annotations. Additionally, these datasets typically focus narrowly on specific diagnostic tasks or cancer types, thereby constraining their broader applicability and limiting model robustness \citep{li2025surveycomputationalpathologyfoundation}.

To accelerate progress and address these limitations, there is a clear need for openly accessible, large-scale, and multimodal datasets in Digital Pathology. Such datasets would enable researchers and developers to explore more generalizable, robust, and clinically relevant AI models capable of real-world deployment.

In this paper, we introduce and open-source the HISTAI-dataset, a substantial WSI dataset designed to bridge the existing gaps by providing a diverse collection of cases accompanied by comprehensive clinical metadata. Each case in our dataset includes Whole Slide Images along with metadata containing diagnostic information (including diagnosis and ICD-10 codes), patient demographics (such as age and gender), detailed pathological conclusions, differential diagnostic considerations, as well as macro and micro protocol descriptions. By releasing this dataset to the community, we aim to foster innovation, promote reproducibility, and ultimately drive the DP industry forward towards more accurate and generalized AI solutions.

\section{Related Work}

A number of public whole-slide image (WSI) datasets have been published throughout the years in digital pathology, particularly for supervised learning and benchmarking. Below, we review the most notable slide-level resources, each with its own motivation, design constraints, and domain focus.

\begin{itemize}

  \item \textbf{TCGA} \cite{tcga_gdc}:  
  The Cancer Genome Atlas (TCGA) was a landmark initiative by the NIH and NCI to map genomic alterations across cancer types. As part of this effort, approximately \textbf{33,500} H\&E-stained diagnostic WSIs from \textbf{11,000} patients were digitized across \textbf{33} tumor types. These slides, while primarily included for diagnostic documentation, became a de facto standard for deep learning in pathology due to their breadth and linkage with transcriptomic, genomic, and survival data.

  \item \textbf{CAMELYON16} \cite{camelyon16}:  
  Developed as part of the CAMELYON16 challenge at ISBI 2016, this dataset includes \textbf{400} WSIs of breast cancer sentinel lymph nodes collected at two Dutch institutions. Its purpose was to benchmark algorithms for detecting metastases in lymph node tissue. Notably, \textbf{129} slides have pixel-wise annotations outlining tumor metastases, enabling precise evaluation of detection performance.

  \item \textbf{CAMELYON17} \cite{camelyon17}:  
  A follow-up to CAMELYON16, this dataset was built for the CAMELYON17 challenge at MICCAI 2017 to promote robust, patient-level prediction. It contains \textbf{1,000} H\&E WSIs from \textbf{200} patients across five centers, each with slide-level and patient-level metastasis labels (e.g., isolated tumor cells, micro- or macro-metastases). Fifty slides are annotated with detailed tumor segmentation masks, while the full set supports evaluation of algorithms under inter-center variability and domain shift.

  \item \textbf{TUPAC16} \cite{tupac16}:  
  Released during the TUPAC16 challenge at MICCAI 2016, this dataset addressed automated assessment of tumor proliferation in breast cancer. It includes \textbf{500} WSIs from TCGA-BRCA, each labeled with a tumor proliferation score derived from mitotic activity and mRNA expression. Unlike CAMELYON, the goal here was slide-level regression/classification rather than segmentation. A separate mitosis detection task was based on ROI patches, making TUPAC one of the first datasets to link WSI-level outcomes to molecular biomarkers.

  \item \textbf{PANDA} \cite{panda2022}:  
  The PANDA dataset was released as part of a Kaggle competition in 2020, motivated by the clinical demand for scalable Gleason grading in prostate biopsies. It comprises \textbf{10,616} H\&E-stained needle core biopsy slides from Radboud UMC and Karolinska Institute. Each slide is labeled with a Gleason grade group (0–5), enabling direct supervised learning. PANDA is notable for its unprecedented size, cross-center diversity, and real-world diagnostic setting, making it a dominant benchmark in prostate pathology.

  \item \textbf{ACROBAT} \cite{acrobat2022}:  
  Designed for the ACROBAT challenge at MICCAI 2022, this dataset tackles the problem of registering H\&E and IHC slides. It includes \textbf{4,212} WSIs from \textbf{1,153} breast cancer patients, with up to five stainings per patient: H\&E, ER, PR, HER2, and Ki-67. Each slide pair is aligned via \textbf{37,208} manually annotated landmark correspondences. ACROBAT is one of the few resources enabling multimodal registration, a key task for combining morphological and molecular information.

\end{itemize}

While these datasets have been foundational, key gaps remain. Many organs and cancer types are underrepresented, IHC and other non-H\&E stains are rarely available at scale, and few datasets provide multimodal annotations suited for generalization studies. Our proposed dataset aims to address several of these limitations by releasing HISTAI-dataset.

\section{Dataset Description}

The HISTAI dataset is a comprehensive, open-source resource for computational pathology research, designed to address the limitations identified in existing public datasets. The dataset provides extensive Whole Slide Images (WSIs) accompanied by detailed clinical and pathological metadata, enabling diverse applications ranging from diagnostic modeling to multimodal analyses.

The HISTAI dataset serves as a foundational resource within the broader research ecosystem. Notably, the SPIDER dataset \citep{nechaev2025spidercomprehensivemultiorgansupervised}, a collection of patch-level annotated pathological datasets, utilizes slides from the HISTAI dataset. Additionally, the Hibou foundation models \citep{nechaev2024hiboufamilyfoundationalvision}, have been trained on a larger, encompassing dataset, of which HISTAI forms a subset.

\subsection{Structure and Organization}

The dataset is organized into specialized subsets based on tissue types and pathology specializations. These subsets are independently accessible on the Hugging Face platform, allowing researchers targeted access to relevant cases. The general naming structure for slides within each case is as follows:

\begin{itemize}
\item \texttt{<dataset\_name>/case\_<case\_id>/slide\_<stain>\_<slide\_number>.tiff}
\item \texttt{<dataset\_name>/case\_<case\_id>/slide\_<magnification>\_<stain>\_<slide\_number>.tiff}
\end{itemize}

Most slides are digitized at a standard magnification of 20X using Hematoxylin and Eosin (H\&E) staining. Slides captured at alternative magnifications (primarily 40X) explicitly note the magnification in their filenames, ensuring clarity for downstream tasks. The majority of the slides were digitized using Leica Aperio GT450 and AT2 scanners, though some were scanned with Hamamatsu or 3DHISTECH systems. However we can't determine the exact model of the scanner for individual slides.

\subsection{Metadata}

A master metadata repository accompanies the image subsets, providing extensive clinical, pathological, and technical annotations for each case. The metadata includes:

\begin{itemize}
\item \textbf{Diagnosis:} Clinical diagnostic information and notes.
\item \textbf{Conclusion:} Detailed pathological conclusions.
\item \textbf{Differential Diagnosis:} Alternative diagnostic considerations.
\item \textbf{Micro Protocol:} Detailed microscopic examination descriptions.
\item \textbf{Additional Information:} Supplementary clinical or pathological details.
\item \textbf{Patient Demographics:} Age and gender of the patient.
\item \textbf{ICD-10 Codes:} Standardized diagnostic codes.
\item \textbf{Gross Examination Details:} Descriptions from macroscopic analysis.
\end{itemize}

Each metadata entry directly references the corresponding WSIs, ensuring seamless integration of clinical and image-based data.

\subsection{Available Specialized Datasets}

Currently, HISTAI comprises the following specialized subsets:

\begin{itemize}
\item \textbf{HISTAI-hematologic}
\item \textbf{HISTAI-gastrointestinal}
\item \textbf{HISTAI-breast}
\item \textbf{HISTAI-thorax}
\item \textbf{HISTAI-skin-b1}
\item \textbf{HISTAI-skin-b2}
\item \textbf{HISTAI-colorectal-b1}
\item \textbf{HISTAI-colorectal-b2}
\end{itemize}

\subsection{Dataset Statistics}

The HISTAI dataset currently contains a substantial number of WSIs and cases, summarized in Table~\ref{table-stats}.

\begin{table}[htb]
\centering
\caption{HISTAI Dataset Statistics}
\label{table-stats}
\begin{tabular}{lrr}
\toprule
\textbf{Dataset} & \textbf{Total Slides} & \textbf{Total Cases} \\
\midrule
HISTAI-hematologic & 214 & 214 \\
HISTAI-gastrointestinal & 202 & 120 \\
HISTAI-breast & 1,925 & 1,692 \\
HISTAI-thorax & 829 & 657 \\
HISTAI-skin-b2 & 43,757 & 20,621 \\
HISTAI-skin-b1 & 7,710 & 1,778 \\
HISTAI-colorectal-b1 & 5,379 & 998 \\
HISTAI-colorectal-b2 & 94 & 62 \\
\midrule
\textbf{Total} & \textbf{60,110} & \textbf{26,142} \\
\bottomrule
\end{tabular}
\end{table}

Further breakdowns of the dataset include:

\begin{itemize}
\item \textbf{Magnification:}
\begin{itemize}
\item Slides at 20X magnification: 57,647
\item Slides at 40X magnification: 2,463
\end{itemize}

\item \textbf{Staining protocols:}
\begin{itemize}
    \item H\&E slides: 58,282
    \item Other staining protocols: 1,828
\end{itemize}

\end{itemize}

\subsection{Intended Uses and Potential Applications}

The extensive variety and detailed annotation of the HISTAI dataset facilitate a wide range of potential research applications, including but not limited to:

\begin{itemize}
\item Development and benchmarking of diagnostic models.
\item Studies on generalization across tissue types and clinical contexts.
\item Investigation into multimodal pathology models integrating clinical metadata.
\item Exploration of domain adaptation and transfer learning in digital pathology.
\end{itemize}

By releasing the HISTAI dataset, we aim to significantly contribute to ongoing research, enhance reproducibility, and encourage the development of robust, clinically applicable AI solutions in digital pathology.

\section{Conclusion}

The HISTAI dataset represents a significant advancement in addressing the current limitations of publicly available WSI datasets in digital pathology. By providing a comprehensive, multimodal, and richly annotated collection of over 60,000 slides across diverse tissue types, HISTAI facilitates a wide array of computational pathology research opportunities. Its open accessibility and detailed metadata not only promote reproducibility but also support the development of more robust, generalizable, and clinically relevant AI solutions, ultimately advancing the broader field of digital pathology.

\bibliographystyle{unsrtnat}
\bibliography{references}

\end{document}